\documentclass[conference, a4paper]{IEEEtran}

\usepackage{pstricks}
\usepackage{graphicx}
\usepackage{psfrag}
\usepackage[cmex10]{amsmath}
\usepackage{amssymb}
\usepackage{color}
\usepackage{fancyhdr}
\definecolor{mygray}{rgb}{0.6,0.6,0.6}

\def\ve#1{{\mathchoice{\mbox{\boldmath$\displaystyle #1$}}%
              {\mbox{\boldmath$\textstyle #1$}}%
              {\mbox{\boldmath$\scriptstyle #1$}}%
              {\mbox{\boldmath$\scriptscriptstyle #1$}}}}
\DeclareSymbolFont{AMSb}{U}{msb}{m}{n}
\DeclareSymbolFontAlphabet{\mathbb}{AMSb}

\def\F{\mathbb{F}}

\def\conc{\otimes}

\def\labeling{\mathcal L}
\def\mean{M}

\def\W{\mathsf{W}}
\def\B{\mathsf{B}}

\begin{document}
\pagestyle{fancy}
\fancyhead[RE,LO]{\color{mygray}Mathis~Seidl, Andreas Schenk, Clemens Stierstorfer, and Johannes~B.~Huber:\quad Multilevel Polar-Coded Modulation\\
                    Submitted to IEEE ISIT 2013}
\title{Multilevel Polar-Coded Modulation}
 \author{
   \IEEEauthorblockN{Mathis~Seidl,~Andreas~Schenk,~Clemens~Stierstorfer, and~Johannes~B.~Huber}
   \IEEEauthorblockA{Institute of Information Transmission, 
     Friedrich-Alexander-Universit{\"a}t Erlangen-N{\"u}rnberg, Germany\\
     Email: {\tt\{seidl,schenk,clemens,jbhuber\}}@LNT.de.}
 }
\maketitle
%
\begin{abstract}
A framework is proposed that allows for a joint description and optimization of both 
binary polar coding and the multilevel coding (MLC) approach for $2^m$-ary digital 
pulse-amplitude modulation (PAM). 
The conceptual equivalence of polar coding and multilevel coding is pointed out in detail. 
Based on a novel characterization of the channel polarization phenomenon, rules for 
the optimal choice of the bit labeling in this coded modulation scheme employing polar 
codes are developed. 
Simulation results for the AWGN channel are included. 
\end{abstract}

\section{Introduction}
\label{sec:intro}
\noindent
Polar codes~\cite{Arikan:09} are known as a low-complexity binary coding scheme that provably approaches the capacity of arbitrary symmetric binary-input discrete memoryless channels (B-DMCs). 
The generalization to $q$-ary channels ($q>2$) has been the subject of various works, cf., e.g. \cite{Sasoglu:09}.
Though, the topic of \emph{polar-coded modulation}, i.e., the combination of $2^m$-ary digital PAM modulation and binary polar codes for increased spectral efficiency, has hardly been addressed so far. 
In \cite{ShinLimYang:12}, a transmission scheme for polar codes with bit-interleaved coded modulation (BICM)~\cite{CaireTB:98} has been proposed, focussing on the interleaver design. 

In this paper, we consider the multilevel coding (MLC) construction~\cite{ImaiH:1977, WachsmannFH:99} for memoryless channels like the AWGN channel (no 
fading). 

It has been observed (e.g.,~\cite{ArikanISIT:11}) that the MLC approach is closely related to that of polar coding on a conceptual level. 
Based on these similarities, we propose a framework that allows us to completely describe both polar coding and $2^m$-ary modulation in a unified context as certain channel transforms. 
This unified description enables us to design optimized constellation-dependent coding schemes for MLC. 

The paper is organized as follows: 
In Sec.~\ref{sec:ch_transf}, the framework for a joint description of polar coding and $2^m$-ary PAM modulation is developed. 
This framework is then used for describing the polar coding construction in Sec.~\ref{sec:polarcodes}, leading to a novel interpretation of the polarization phenomenon. 
The optimum combination of binary polar coding and multilevel coding is discussed in Sec.~\ref{sec:polar_mlc}, followed by simulation results for the AWGN channel in Sec.~\ref{sec:sim}.


\section{Channel Transforms}
\label{sec:ch_transf}
\noindent
\subsection{Sequential Binary Partitions}
Let $\W : \mathcal X \rightarrow \mathcal Y$ be a discrete, memoryless channel (DMC) with input symbols $x\in \mathcal X$ (alphabet size $| \mathcal X|= 2^k$), output symbols 
$y\in \mathcal Y$ from an arbitrary alphabet $\mathcal Y$, 
and mutual information $I(X;Y)$. 
\footnote{A short remark on the notation: Channels are denoted by sans serif fonts, capital roman letters stand for random variables while boldfaced symbols denote vectors or matrices.}
We define an order-$k$ \emph{sequential binary partition} ($k$-SBP) $\varphi$ of $\W$ to be a channel transform\vspace*{-1mm}
\begin{equation}
 \varphi : \W \rightarrow \{\B_\varphi^{(0)},\ldots, \B_\varphi^{(k-1)} \}
\end{equation}
that maps $\W$ to an \emph{ordered} set of $k$ binary-input 
DMCs (B-DMCs) which we will refer to as \emph{bit channels}. 
For any given $\W$, such a $k$-SBP is characterized  by  a binary labeling rule $\labeling_\varphi$ that maps binary $k$-tuples bijectively to 
the $2^k$ input symbols $x \in \mathcal X$:\vspace*{-1mm}
\begin{equation}
 \labeling_\varphi :\ [b_0,b_1,\ldots ,b_{k-1}] \in \{0,1\}^k \mapsto x \in \mathcal X\; .
\end{equation}
The number of possible labelings equals $(2^k!)$.

Each bit channel $\B_\varphi^{(i)}$ ($0 \leq i < k$) of a $k$-SBP is supposed to have knowledge of the output of $\W$ as well as of the values transmitted over the bit channels of smaller 
indices $\B_\varphi^{(0)},\ldots ,\B_\varphi^{(i-1)}$. Thus, we have\vspace*{-1mm}
\begin{equation}
 \B_\varphi^{(i)}    : \{0,1\} \rightarrow \mathcal Y \times \{0,1\}^i \quad , \quad 0 \leq i < k \; .
\end{equation}
The mutual information between channel input and output of $\B_\varphi^{(i)}$ assuming equiprobable input symbols is therefore given by \vspace*{-1mm}
\begin{equation}
 I(\B_\varphi^{(i)}) := I(B_i ; Y | B_0 , \ldots , B_{i-1})
\end{equation}
which we will refer to as the \emph{(symmetric) bit channel capacity} of $\B_\varphi^{(i)}$. 
(If $\W$ is a symmetric channel, this value in fact equals the channel capacity.)
The mutual information of $\W$ is preserved under the transform $\varphi$, i.e.,\vspace*{-1mm}
\begin{equation}\label{chain_rule}
 \sum_{i=0}^{k-1}I(\B_\varphi^{(i)}) = I(X;Y)
\end{equation}
which directly follows from the well-known chain rule of mutual information.

Considering polar-coded modulation, we show that the code construction can be described by SBPs. We are particularly interested in two properties of SBPs, namely the mean value 
and the variance of the bit channel capacities, defined respectively as\vspace*{-1mm}
\begin{align}
 \mean_\varphi(\W) &:= \frac{1}{k}\sum_{i=0}^{k-1} I(\B_\varphi^{(i)}) = \frac{1}{k} I(X;Y)\label{def_mean}\\ 
 V_\varphi(\W) &:= \frac{1}{k}\sum_{i=0}^{k-1} I(\B_\varphi^{(i)})^2 - \mean_\varphi(\W)^2 \; .\label{def_variance}
\end{align}
Clearly, from \eqref{chain_rule} the mean value $\mean_\varphi(\W)$ in fact depends only on the channel $\W$, rather than on the transform $\varphi$. 
It represents the average (symmetric) capacity of $\W$ per transmitted binary symbol. 

The variance of an SBP $\varphi$ is upper-bounded by \vspace*{-1mm}
\begin{equation}
 V_\varphi(\W) \leq \mean_\varphi(\W) (1- \mean_\varphi(\W))
\end{equation}
with equality only iff all $I(\B_\varphi^{(i)})$ are either $0$ or $1$. 
This follows from\vspace*{-1mm} 
\begin{align}\label{max_var}
 V_\varphi(\W) &=   \frac{1}{k}\sum_{i=0}^{k-1} I(\B_\varphi^{(i)})^2 - \mean_\varphi(\W)^2 \\\nonumber
               &\leq \frac{1}{k}\sum_{i=0}^{k-1} I(\B_\varphi^{(i)}) - \mean_\varphi(\W)^2  \\\nonumber
               &= \mean_\varphi(\W) (1- \mean_\varphi(\W)) \; .
\end{align}
and $0 \leq I(\B_\varphi^{(i)}) \leq 1$ for all $0 \leq i < k$.
Note that this upper bound does not depend on the particular labeling $\labeling_\varphi$ but only on the channel $\W$.

An important subset of $k$-SBPs is formed by those transforms whose labeling rules are described by binary bijective linear mappings. 
Let $\W = (\B_0 \times \ldots \times \B_{k-1})$ be a vector channel of $k$ independent arbitrary B-DMCs $\B_0, \ldots , \B_{k-1}$. 
Then, we call the $k$-SBP\vspace*{-1mm}
\begin{equation}
 \varphi: \quad  (\B_0 \times \ldots \times \B_{k-1}) \rightarrow \{\B_\varphi^{(0)},\ldots, \B_\varphi^{(k_1-1)} \}
\end{equation}
a \emph{linear $k$-SBP} if its labeling rule is given by\vspace*{-1mm}
\begin{equation}
 \labeling_\varphi :\ \ve{b} \in \F_2^k \mapsto \ve{b} \cdot \ve{A}_{\varphi} \in \F_2^k \; .
\end{equation}
with $\ve{b} := [b_0,b_1,\ldots ,b_{k-1}]$ and $\ve{A}_{\varphi}$ being an invertible binary $(k,k)$ matrix. 
Clearly, the number of possible linear $k$-SBPs equals the number of non-singular binary $(k,k)$ matrices and is significantly smaller than that of 
general $k$-SBPs.

\subsection{Product Concatenation of SBPs}
Under certain conditions, it is possible to concatenate two (or more) SBPs in a product form. Let\vspace*{-1mm}
\begin{equation}
 \varphi: \quad \W \rightarrow \{\B_\varphi^{(0)},\ldots, \B_\varphi^{(k_1-1)} \}
\end{equation}
be an arbitrary $k_1$-SBP and \vspace*{-1mm}
\begin{equation}
 \psi: \quad (\B_0 \times \ldots \times \B_{k_2-1}) \rightarrow \{\B_\psi^{(0)},\ldots, \B_\psi^{(k_2-1)} \}
\end{equation}
a $k_2$-SBP that takes a vector channel of $k_2$ independent B-DMCs $\B_0, \ldots , \B_{k_2-1}$ as an input. 
Each of the vector channels $(\B_\varphi^{(i)})^{k_2}$ -- obtained by taking $k_2$ independent instances of $\B_\varphi^{(i)}$ -- can be partitioned by $\psi$.
Thus, $\varphi$ and $\psi$ may be concatenated by considering the vector channel $\W^{k_2}$, leading to a product SBP of order $k_1k_2$:\vspace*{-1mm}
\begin{equation}\label{def_SBP_conc}
 \varphi \conc \psi: \W^{k_2} \rightarrow \{\B_{\varphi \conc \psi}^{(0)},\ldots, \B_{\varphi \conc \psi}^{(k_1k_2-1)} \} \; .
\end{equation}
Here, the bit channels of $\varphi \conc \psi$ are given by\vspace*{-1mm}
\begin{equation}
 \B_{\varphi \conc \psi}^{(k_2i+j)} : \{0,1\} \rightarrow \mathcal Y^{k_2} \times \{0,1\}^{k_2i+j}
\end{equation}
with symmetric capacities\vspace*{-1mm}
\begin{equation}
 I(\B_{\varphi \conc \psi}^{(k_2i+j)}) = I(B_{k_2i+j} ; Y_0,\ldots ,Y_{k_2-1} | B_0 , \ldots , B_{k_2i+j-1})
\end{equation}
such that\vspace*{-1mm}
\begin{equation}\label{mean_conc}
 \frac{1}{k_2}\sum_{j=0}^{k_2-1} I(\B_{\varphi \conc \psi}^{(k_2i+j)}) = I(\B_\varphi^{(i)})
\end{equation}
for all $0\leq i < k_1$ and $0\leq j < k_2$. We remark that the product transform $\varphi \conc \psi$ is completely determined in a unique way
by the individual SBPs $\varphi$ and $\psi$ since their bit channels imply a fixed order. 

The product concatenation of SBPs does not influence the mean value of the bit channel capacities\vspace*{-1mm}
\begin{align}\label{mean_invar}
 \mean_{\varphi \conc \psi}(\W^{k_2}) &= \mean_{\varphi}(\W)
\end{align}
due to the chain rule of mutual information. 
However, the variance of the bit channel capacities increases. It is given by the sum of the variance of the first transform and the averaged variance of the second transform around 
the bit channel capacities of the first one:\vspace*{-1mm}
\begin{equation}\label{var_conc}
 V_{\varphi \conc \psi}(\W^{k_2}) = V_{\varphi}(\W) + \frac{1}{k_1} \sum_{i=0}^{k_1-1} V_\psi(\B_\varphi^{(i)}) \; .
\end{equation}

If $\varphi$ and $\psi$ are linear SBPs with labeling rules specified by $\ve{A}_\varphi$ and $\ve{A}_\psi$, respectively, then their product $\varphi \conc \psi$ is again a linear 
$k_1k_2$-SBP with labeling rule\vspace*{-1mm}
\begin{equation}\label{SBP_conc_lin}
 \labeling_{\varphi \conc \psi} :\ \ve{b} \in \F_2^{k_1k_2} \mapsto \ve{b} \cdot \ve{P}_{k_1,k_2} \cdot \left( \ve{A}_{\psi} \otimes \ve{A}_{\varphi} \right) \; .
\end{equation}
Here, $ \ve{A}_{\psi} \otimes \ve{A}_{\varphi}$ denotes the Kronecker product of $\ve{A}_{\psi}$ and $\ve{A}_{\varphi}$. 
$\ve{P}_{k_1,k_2}$ is the $(k_1k_2,k_1k_2)$ permutation matrix that maps the ($k_2i+j$)-th component of the vector $\ve{b}$ to position $i+k_1j$ 
(for all $0\leq i < k_1$, $0 \leq j < k_2$).


\section{Polar Codes}
\label{sec:polarcodes}
\noindent

Polar codes, as introduced by Ar{\i}kan~\cite{Arikan:09}, have been shown to be a channel coding construction that provably achieves the symmetric capacity of arbitrary binary-input discrete memoryless channels (B-DMCs) under low-complexity encoding and successive cancellation (SC) decoding. 
For sake of simplicity, we focus on Ar{\i}kan's original construction in this paper; the generalization to polar codes based on different kernels (as considered, e.g., in~\cite{Korada:10}) 
is straightforward. Furthermore, we restrict our considerations to the SC decoding algorithm as in~\cite{Arikan:09}; though, our results regarding the code construction are also 
valid for other (better performing) decoders that are based on the SC algorithm, as, e.g., list decoding~\cite{TalVardy:11}.

\subsection{Code Construction}

Let $\B : \{0,1\} \rightarrow \mathcal Y$ be a B-DMC and $I(\B)$ its symmetric capacity, i.e., the mutual information of $\B$ assuming equiprobable binary input symbols. 
The encoding operation for a polar code of length $N$ may be 
described by multiplication of a binary length-$N$ vector $\ve{u}$ -- containing the information symbols as well as some symbols with fixed values (so-called \emph{frozen symbols}) 
that do not carry any information -- 
with a generator matrix $\ve{G}_N$ that is defined by the 
recursive relation\vspace*{-1mm}
\begin{equation}
 \ve{G}_N = \ve{B}_N\ve{F}_N\; , \quad \ve{F}_{2N} = \ve{F}_2 \otimes \ve{F}_N \; , \quad  \ve{F}_2 = \left[ \begin{matrix} 1 & 0\\ 1 & 1 \end{matrix} \right]  \label{eq:polarc:G}
\end{equation}
where $N$ is a power of two and $\otimes$ again denotes the Kronecker product. $\ve{B}_N$ denotes the $(N,N)$ bit-reversal permutation matrix~\cite{Arikan:09}. 
Encoding takes place in the binary field $\F_2$. 
The resulting codeword $\ve{c}=\ve{u}\ve{G}_N$ is then transmitted in $N$ time steps over the channel $\B$. 

The code construction is based on a channel combining and channel splitting operation~\cite{Arikan:09} that may be represented as a linear $2$-SBP\vspace*{-1mm}
\begin{equation}
 \pi : \B^2 \rightarrow \{\B_\pi^{(0)}, \B_\pi^{(1)} \}
\end{equation}
that partitions the vector channel $\B^2$, i.e., two independent and identical instances of $\B$, into two bit channels\vspace*{-1mm}
\begin{align}
 \B_\pi^{(0)}    &: \{0,1\} \rightarrow \mathcal Y^2 \\\nonumber
 \B_\pi^{(1)}    &: \{0,1\} \rightarrow \mathcal Y^2 \times \{0,1\} \;.
\end{align}
The labeling rule is given by\vspace*{-2mm}
\begin{equation}
\labeling_\pi : \ \ve{u} = [u_0,u_1] \in \{0,1\}^2 \mapsto \ve{u} \cdot \ve{G}_2 \in \{0,1\}^2 \; .
\end{equation}
Since the average capacity per binary symbol does not change under an SBP, we denote the mean value of the bit channel capacities of $\pi$ by $I(\B)$ instead of $\mean_\pi(\B)$ in the following. 

The construction of a polar code of length $N=2^n$ may be equivalently represented by the $n$-fold product concatenation of $\pi$ as defined in the preceding section. 
This follows easily from~\cite{Arikan:09} by comparison of the corresponding permutation matrices. 
The resulting SBP $\pi^n$ partitions the vector channel $\B^N$\vspace*{-1mm}
\begin{equation}
 \pi^n : \B^N \rightarrow \{\B_{\pi^n}^{(0)}, \ldots, \B_{\pi^n}^{(N-1)} \}
\end{equation}
into $N$ bit channels\vspace*{-1mm}
\begin{equation}
 \B_{\pi^n}^{(i)}    : \{0,1\} \rightarrow \mathcal Y^N \times \{0,1\}^i
\end{equation}
($0\leq i< N$) with symmetric capacities\vspace*{-1mm}
\begin{equation}
 I(\B_{\pi^n}^{(i)}) := I(U_i ; Y_0,\ldots ,Y_{N-1} | U_0 , \ldots , U_{i-1}) \; .
\end{equation}

Therefore, the transmission of each source symbol $u_i$ can be described by its own bit channel $\B_{\pi^n}^{(i)}$.
The output of each channel $\B_{\pi^n}^{(i)}$ depends on the values of the symbols of lower indices $u_0,\ldots u_{i-1}$. 
Thus, the channels $\B_{\pi^n}^{(i)}$ imply a specific decoding order.

For data transmission only the bit channels with highest capacity
are used, referred to as \emph{information channels}. 
The data transmitted over the remaining bit channels (so-called \emph{frozen channels})
are fixed values known to the decoder. 
By this means, the code rate can be chosen in very small steps of $1/N$ without the need for changing the code construction -- 
a property especially useful for polar-coded modulation (cf., Sec.~\ref{sec:polar_mlc}-B).

In order to select the optimal set of frozen channels, the values of
the capacities $I(\B_{\pi^n}^{(i)})$ are required. These 
can either be obtained by simulation or by density evolution~\cite{Mori:09}.

\subsection{Successive Decoding}
Upon receiving a vector $\ve{y}$ --  being a noisy version of the codeword $\ve{c}$ resulting from transmission over the channel $\B$ -- 
the information bits $u_i$ can be estimated successively for $i=0,\ldots, N-1$. 
Here, information combining~\cite{LandH:2006} of reliability values obtained from the channel output $\ve{y}$ is performed instead of $\F_2$ arithmetics as in the encoding process. 

The successive cancellation (SC) decoding algorithm~\cite{Arikan:09} for polar codes generates estimates on the information symbols $\hat u_i$ (transmitted over $\B_{\pi^n}^{(i)}$) 
one after another, making use of the already decoded symbols $\hat u_0,\ldots,\hat u_{i-1}$. 
We denote the probability that an erroneous decision is made at index $i$ given the previous decisions have been correct, by $p_{\mathrm{e}}(\B_{\pi^n}^{(i)})$. 
Thus, the word error rate for SC decoding ($\mathrm{WER}_{\mathrm{SC}}$) is given by\vspace*{-1mm}
\begin{equation}\label{sc_wer}
 \mathrm{WER}_{\mathrm{SC}} = 1 - \prod_{i \in \mathcal A}\left(1 - p_{\mathrm{e}}(\B_{\pi^n}^{(i)})\right)
\end{equation}
where $\mathcal A$ denotes the set of indices of the information channels.

\subsection{Variance of the Bit Channel Capacities}
With increasing block length, the set of bit channels $\B_{\pi^n}^{(i)}$ shows a polarization
effect in the sense that the capacity $I(\B_{\pi^n}^{(i)})$ of almost each bit channel is
either near $0$ or near $1$. The fraction of bit channels not being either completely noisy or
completely noiseless tends to zero \cite{Arikan:09}. 

In the following, we show that 
this polarization effect may be represented by the sequence of variances of the respective polar codes' bit channel capacities for increasing block length.
The variance of the bit channel capacities of a length-$N$ polar code around their mean value $I(\B)$ is given by\vspace*{-1mm}
\begin{equation}
 V_{\pi^n}(\B^N) = \frac{1}{N}\sum_{i=0}^{N-1} I(\B_{\pi^n}^{(i)})^2 - I(\B)^2 \; .
\end{equation}
Using \eqref{var_conc}, we notice that the sequence of variances increases monotonously as the block length gets larger, i.e.,\vspace*{-1mm}
\begin{equation}
 V_{\pi^{n+1}}(\B^{2N}) \geq V_{\pi^n}(\B^N) \; .
\end{equation}
Furthermore, from \eqref{max_var} the sequence $\{V_{\pi^n}(\B^N)\}_{n \in \mathbb N}$ is upper-bounded by\vspace*{-1mm}
\begin{equation}
 V_{\pi^n}(\B^N) \leq I(\B)(1-I(\B))
\end{equation}
for all $n \in \mathbb N$. According to \eqref{max_var}, this maximum variance can be only achieved iff all bit channel capacities $I(\B_{\pi^n}^{(i)})$ are either $0$ or $1$, which 
obviously corresponds to the state of perfect polarization. As shown by Ar{\i}kan~\cite{Arikan:09}, the latter is asymptotically approached while the block length $N$ goes to infinity; 
therefore, we have\vspace*{-1mm}
\begin{equation}
 \lim_{n \rightarrow \infty} V_{\pi^n}(\B^N) =  I(\B) \cdot (1-I(\B)) \; .
\end{equation}
Although we have not yet been able to establish an explicit relation between bit channel capacity variance and code error performance, one would intuitively expect 
that increasing the variance by a careful code design should correspond to a sharper polarization of the bit channels and therefore should lead to better performing polar 
codes in terms of word error rate or bit error rate.

\begin{figure}
 \psfrag{Clabel}[Bc][Bc][0.8]{$I(\B_\mathrm{BEC})$}
 \psfrag{Vlabel}[Bc][Bc][0.8]{$V_{\pi^n}(\B_\mathrm{BEC}^N)$}
 \centerline{\includegraphics[width=0.45\textwidth]{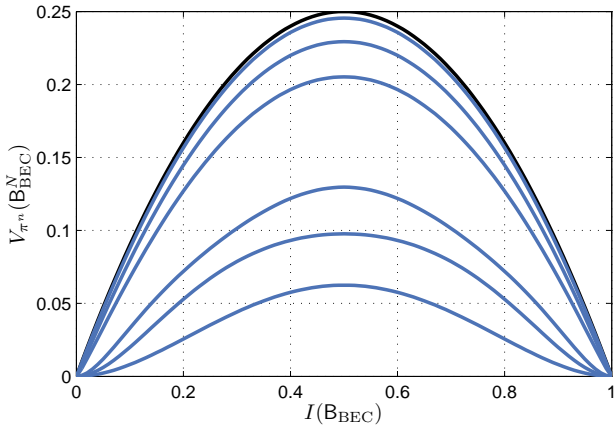}}%
 \vspace*{-3mm}\caption{\label{var_bec}Bit channel variance for polar codes over a BEC channel $\B_\mathrm{BEC}$, block length $N=2^n$, $n=1,2,3,8,12,20$. 
 Black: upper bound on the variance.}
 \vspace*{-4mm}
\end{figure}

Fig.~\ref{var_bec} depicts the variance of the bit channel capacities for polar codes of various block lengths constructed over the BEC channel 
as a function of its capacity. 
The converging behaviour for increasing block length $N$ towards the maximum achievable variance (black line) 
can clearly be observed.


\section{Multilevel Polar Coding}
\label{sec:polar_mlc}
\noindent
We now consider the conventional discrete-time equivalent system model of $M$-ary digital pulse-amplitude modulation (PAM)  -- $M=2^m$ being a power of $2$ --
with signal constellations of real-valued signal points (ASK) or of complex-valued signal points (PSK, QAM etc.) over 
a memoryless channel $\W$, e.g., the AWGN channel. 

From an information-theoretic point of view, an optimum combination of binary coding and $M$-ary modulation follows the multilevel coding (MLC) principle \cite{ImaiH:1977, WachsmannFH:99}.

\subsection{Multilevel Coding}
In the MLC approach, the $M$-ary channel $\W$ is partitioned into $m$ bit channels (also called \emph{bit levels}) by means of an $m$-SBP\vspace*{-1mm}
\begin{equation}\label{mlc_labeling}
 \lambda : \W \rightarrow  \{\B_\lambda^{(0)},\ldots, \B_\lambda^{(m-1)} \} \; .
\end{equation}
The mapping from binary labels to amplitude coefficients is specified by the labeling rule $\labeling_\lambda$.

Channel coding is implemented in the MLC setup by using binary \emph{component codes} \cite{WachsmannFH:99} for each of the bit levels $\B_\lambda^{(i)}$ individually 
with correspondingly chosen code rates $R_i$. The overall rate (bits per transmission symbol) is given as the sum $R=\sum_{i=0}^{m-1}R_i$. 
The receiver then performs multi-stage decoding (MSD), i.e., it computes reliability information for decoding of the first bit level which are passed to the decoder of 
the first component code. The decoding results are used for demapping and decoding of the next bit level, and so on.

The mutual information between the channel input and channel output of $\W$ assuming equiprobable source symbols is also referred to as the 
coded modulation~\cite{WachsmannFH:99}, or constellation-constraint, capacity $C_{\mathrm{cm}}(\W)$. It is related to the average capacity per binary symbol \eqref{def_mean} 
of $\W$  by\vspace*{-1mm}
\begin{align}\label{eq:cm_cap}
   C_{\mathrm{cm}}(\W)  :=  I(X;Y) =  \sum_{i=0}^{m-1}I(\B_\lambda^{(i)}) = m \cdot \mean_\lambda(\W) \; .
\end{align}
Since $\lambda$ is an SBP, the coded modulation capacity does not depend on the specific labeling rule $\labeling_\lambda$. 

A potential drawback of the MLC approach for practical use lies in the necessity for using several (relatively short) component codes with varying code rates for the particular bit levels. 
According to the capacity rule~\cite{WachsmannFH:99}, the code rate $R_i$ for the $i$-th level should match the bit level capacity $R_i = I(\B_\lambda^{(i)})$. Since these capacities vary significantly for the different levels, for MLC channel codes are preferred, that allow for a very flexible choice of the code rate.

\subsection{Multilevel Polar Coding}
A \emph{multilevel polar code} of length $mN$, i.e., a multilevel code with length-$N$ component polar codes over an $M$-ary constellation, is obtained by the 
order-$mN$ concatenation of the $m$-SBP $\lambda$ of MLC and the $N$-SBP $\pi^n$ of the polar code: \vspace*{-1mm}
\begin{equation}
 \lambda \conc \pi^n: \W^{N} \rightarrow \{\B_{\lambda \conc \pi^n}^{(0)},\ldots, \B_{\lambda \conc \pi^n}^{(mN-1)} \}
\end{equation}
as defined in \eqref{def_SBP_conc}. 
The encoding process for this multilevel polar code is described by the generator matrix\vspace*{-1mm}
\begin{equation}
 \ve{P}_{m,N} \cdot \left( \ve{G}_N \otimes \ve{I}_m \right)
\end{equation}
with $\ve{P}_{m,N}$ as in \eqref{SBP_conc_lin}, followed by labeling and mapping to the $N$ transmit symbols as defined by $\lambda$. Here, $\ve{I}_m$ denotes the $(m,m)$ identity matrix. 

We remark that the selection of frozen channels -- and thus, the rate allocation -- is done in exactly the same way as for a usual binary polar code by determining the symmetric capacities 
$I(\B^{(i)})$ ($0\leq i < mN$) and choosing the most reliable bit channels for data transmission. 
Therefore, the explicit application of a rate allocation rule to the particular component codes 
-- like considered in the original MLC approach~\cite{WachsmannFH:99} -- is not needed in case of multilevel polar codes. 
However, it has been shown~\cite{LIT_scc13_seidl_A} that the rate allocations obtained by this method basically equal those obtained from the capacity rule. 

According to \eqref{var_conc}, the variance of the bit channels of a multilevel polar code with length-$N$ component codes is given by\vspace*{-1mm}
\begin{equation}\label{var_mlc_polar}
 V_{\lambda\conc\pi^n}(\W) = V_\lambda(\W) + \frac{1}{m}\sum_{i=0}^{m-1}V_{\pi^n}(\B_\lambda^{(i)}) \; .
\end{equation}
Thus, the SBP $\lambda$ -- that represents the modulation step -- may be seen as the first polarization step of a multilevel polar code. 
From this representation, it is clear that $\lambda$ should be chosen such that it maximizes the term \eqref{var_mlc_polar}. 

In this approach, both binary coding and $2^m$-ary modulation are represented in a unified form as a sequential binary channel partition of the vector channel $\W^N$. Both 
should be designed according to the polarization principle, i.e. the maximization of the variance of the bit channel capacities \eqref{var_mlc_polar} under successive cancellation 
-- or, equivalently, multi-stage -- decoding by careful choice of the labeling rule.

\subsection{Influence of the Labeling Rule}
Here, we focus on the set-partitioning labeling approach (corresponding to $\lambda_{\mathrm{SP}}$) by Ungerboeck~\cite{Ungerboeck:87} 
and Gray labeling $\lambda_{\mathrm{G}}$ that aims to generate bit levels that are as independent as possible~\cite{StierstorferF:vtc07}. 

\begin{figure}
 \psfrag{Clabel}[Bc][Bc][0.8]{$\mean_\lambda(\W)$}
 \psfrag{Vlabel}[Bc][Bc][0.8]{$V_\lambda(\W)$}
 \centerline{\includegraphics[width=0.45\textwidth]{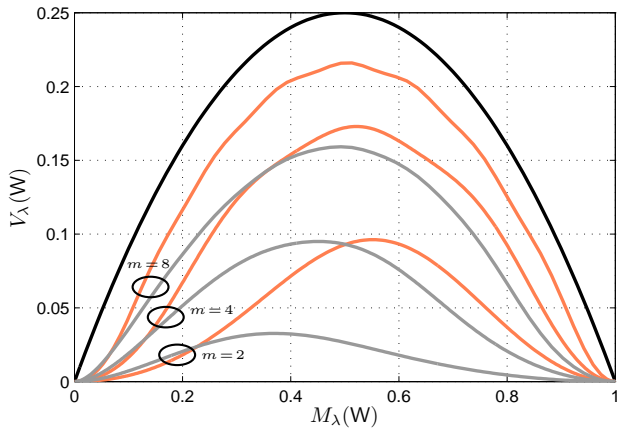}\psellipse(-5.9,1)(.25,.15)\psellipse(-6.05,1.5)(.25,.15)\psellipse(-6.25,1.9)(.25,.15)%
  \rput(-5.3,1){\tiny $m\!=\!2$}\rput(-5.45,1.6){\tiny $m\!=\!4$}\rput(-6.28,2.2){\tiny $m\!=\!8$}}
 \vspace*{-3mm}\caption{\label{var_ask}Bit level variance for $2^m$-ary ASK signalling with multi-stage decoding over the AWGN channel ($m = 2,4,8$). 
  Red: SP labeling, gray: Gray labeling.}
 \vspace*{-4mm}
\end{figure}

Fig. \ref{var_ask} depicts the variance of the bit levels for ASK modulation using both SP and (binary-reflected) Gray labeling. 
It can be observed that -- except for small capacities $\mean_\lambda(\W)$ -- the SP labeling approach leads to significantly larger bit level variances compared to the Gray labeling, as expected. 
Therefore, for multilevel polar codes, SP labeling should be preferably applied.
Furthermore, when compared to the corresponding variance curves of polar codes over a single B-DMC for $N=2,4,8$ as shown in Fig.~\ref{var_bec},
especially in case of SP labeling the achieved bit level variance is significantly higher, emphasizing the importance of the careful choice of the labeling $\labeling_\lambda$ in this 
first step of polarization for multilevel polar codes. 


\section{Simulation Results}
\label{sec:sim}
\noindent
We finally present some numerical results in terms of rate-vs.-power-efficiency plots in order to illustrate the error performance of polar-coded modulation with SC decoding over the AWGN channel.

Besides common Monte-Carlo simulations, we also present approximate results obtained by density evolution (DE)~\cite{Mori:09}.
Here, for multilevel polar codes, we numerically determine the bit channel capacities $I(\B_\lambda^{(i)})$ of the transform $\lambda$ as defined in \eqref{mlc_labeling}.
Then, we calculate the bit channel capacities -- and the corresponding error probabilities $p_e(\B_{\lambda \circ \pi^n}^{(i)})$ -- of the component polar codes 
by performing density evolution with the well-known Gaussian approximation~\cite{ChungUrbanke:01}, i.e., we simply assume the output bit channels of each SBP in the chain 
$\lambda \conc \pi \conc \ldots \conc \pi$ to be Gaussian. 
The word error rate under SC decoding $\mathrm{WER}_\mathrm{SC}$ is obtained from \eqref{sc_wer}.

\begin{figure}
 \psfrag{ebno}[Bc][Bc][0.8]{$10\log_{10}(E_\mathrm{b}/N_\mathrm{0})$}
 \psfrag{rate}[Bc][Bc][0.8]{$R$}
 \centerline{\includegraphics[width=0.42\textwidth]{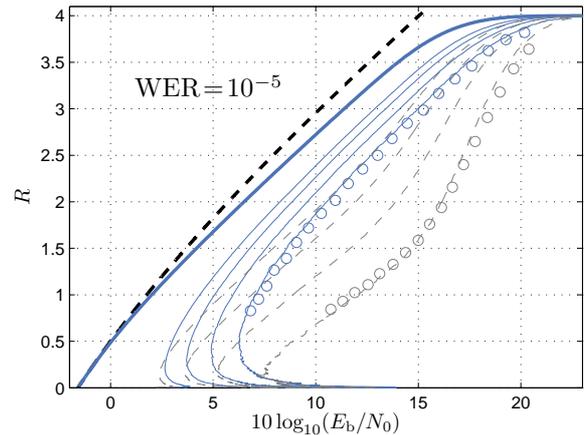}\rput(-5,4.65){$\mathrm{WER}\!=\!10^{-5}$}}
 \vspace*{-3mm}\caption{\label{sim_16ask}$16$-ASK / AWGN: Rate vs. power efficiency of multilevel polar codes using SP (blue) and Gray labeling (dashed gray). 
 Markers: Simulation points for $mN=512$, lines: DE results with
 overall block length (from right to left) $mN=2^k$, 
 $k=9,11,13,15$. Bold blue: coded-modulation capacity, dashed black: Shannon bound for real constellations.}
 \vspace*{-4mm}
\end{figure} 
Fig.~\ref{sim_16ask} depicts the performance of multilevel polar codes with $16$-ASK modulation under SC decoding for different labelings $\labeling_\lambda$ and various block lengths. 
The large performance loss of Gray labeling w.r.t. SP labeling can clearly be observed. Furthermore, by comparison of the results obtained by DE and the simulation points, 
the inaccuracy induced by the Gaussian assumption can be obviously neglected.


\section{Conclusions}
\label{sec:conclusions}
\noindent
In this paper, we have combined polar coding and multilevel coding by representing both 
as sequential binary partitions (SBPs). Based on this representation, we have derived rules for 
optimization of multilevel polar codes. 

Future work will extend this novel framework of channel partitions to bit-interleaved coded modulation (BICM) and 
incorporate fading scenarios. 


\bibliographystyle{IEEEtran}
\vspace*{1mm}
\bibliography{mjssBib}

\begin{thebibliography}{10}
\providecommand{\url}[1]{#1}
\csname url@samestyle\endcsname
\providecommand{\newblock}{\relax}
\providecommand{\bibinfo}[2]{#2}
\providecommand{\BIBentrySTDinterwordspacing}{\spaceskip=0pt\relax}
\providecommand{\BIBentryALTinterwordstretchfactor}{4}
\providecommand{\BIBentryALTinterwordspacing}{\spaceskip=\fontdimen2\font plus
\BIBentryALTinterwordstretchfactor\fontdimen3\font minus
  \fontdimen4\font\relax}
\providecommand{\BIBforeignlanguage}[2]{{%
\expandafter\ifx\csname l@#1\endcsname\relax
\typeout{** WARNING: IEEEtran.bst: No hyphenation pattern has been}%
\typeout{** loaded for the language `#1'. Using the pattern for}%
\typeout{** the default language instead.}%
\else
\language=\csname l@#1\endcsname
\fi
#2}}
\providecommand{\BIBdecl}{\relax}
\BIBdecl

\bibitem{Arikan:09}
E.~Ar{\i}kan, ``Channel polarization: A method for constructing
  capacity-achieving codes for symmetric binary-input memoryless channels,''
  \emph{IEEE Trans. Inf. Theory}, vol.~55, pp. 3051--3073, Jul. 2009.

\bibitem{Sasoglu:09}
E.~Sasoglu, E.~Telatar, and E.~Ar{\i}kan, ``Polarization for arbitrary discrete
  memoryless channels,'' in \emph{Proc. IEEE Inf. Theory Workshop (ITW)}, Oct.
  2009, pp. 144--148.

\bibitem{ShinLimYang:12}
D.-M. Shin, S.-C. Lim, and K.~Yang, ``Mapping selection and code construction
  for 2m-ary polar-coded modulation,'' \emph{IEEE Commun. Lett.}, vol.~16, pp.
  905--908, Jun. 2012.

\bibitem{CaireTB:98}
G.~Caire, G.~Taricco, and E.~Biglieri, ``{Bit-interleaved coded modulation},''
  \emph{IEEE Trans. Inf. Theory}, vol.~44, pp. 927--946, May 1998.

\bibitem{ImaiH:1977}
H.~Imai and S.~Hirakawa, ``A new multilevel coding method using
  error-correcting codes,'' \emph{IEEE Trans. Inf. Theory}, vol.~23, pp.
  371--377, 1977.

\bibitem{WachsmannFH:99}
U.~Wachsmann, R.~F.~H. Fischer, and J.~B. Huber, ``{Multilevel codes:
  Theoretical concepts and practical design rules},'' \emph{IEEE Trans. Inf.
  Theory}, vol.~45, pp. 1361--1391, Jul. 1999.

\bibitem{ArikanISIT:11}
E.~Ar{\i}kan, ``Polar coding: Status and prospects,'' Plenary Talk at
  \emph{IEEE Int. Symp. Inf. Theory (ISIT)}, Aug. 2011, available:
  \url{http://polaran.com/images/arikan2011isit.pdf}.

\bibitem{Korada:10}
S.~Korada, E.~Sasoglu, and R.~Urbanke, ``Polar codes: Characterization of
  exponent, bounds, and constructions,'' \emph{IEEE Trans. Inf. Theory},
  vol.~56, pp. 6253--6264, Dec. 2010.

\bibitem{TalVardy:11}
I.~Tal and A.~Vardy, ``List decoding of polar codes,'' in \emph{Proc. IEEE Int.
  Symp. Inf. Theory (ISIT)}, Aug. 2011, pp. 1--5.

\bibitem{Mori:09}
R.~Mori and T.~Tanaka, ``Performance of polar codes with the construction using
  density evolution,'' \emph{IEEE Commun. Lett.}, vol.~13, pp. 519--521, Jul.
  2009.

\bibitem{LandH:2006}
I.~Land and J.~B. Huber, ``Information combining,'' \emph{Found. Trends Commun.
  Inf. Theory}, vol.~3, no.~3, pp. 227--330, Nov. 2006.

\bibitem{LIT_scc13_seidl_A}
M.~Seidl, A.~Schenk, C.~Stierstorfer, and J.~B. Huber, ``{Aspects of
  Polar-Coded Modulation},'' in \emph{Proc. 9th Int. ITG Conf. Systems, Commun.
  and Coding (SCC)}, Munich, Germany, Jan. 2013.

\bibitem{Ungerboeck:87}
G.~Ungerboeck, ``{Trellis-coded modulation with redundant signal sets Part I:
  Introduction},'' \emph{IEEE Commun. Mag.}, vol.~25, no.~2, pp. 5--11, Feb.
  1987.

\bibitem{StierstorferF:vtc07}
C.~Stierstorfer and R.~F.~H. Fischer, ``{(Gray)} mappings for bit-interleaved
  coded modulation,'' in \emph{Proc. IEEE Veh. Technol. Conf. Spring (VTC
  Spring)}, Dublin, Ireland, Apr. 2007, pp. 1703--1707.

\bibitem{ChungUrbanke:01}
S.-Y. Chung, T.~Richardson, and R.~Urbanke, ``Analysis of sum-product decoding
  of low-density parity-check codes using a gaussian approximation,''
  \emph{IEEE Trans. Inf. Theory}, vol.~47, pp. 657--670, Feb. 2001.

\end{thebibliography}

%
\end{document}